\documentclass[twocolumn,twoside,slac_two,nofootinbib]{revtex4}
\usepackage{graphicx}
\usepackage{fancyhdr}
\usepackage{amsmath}
\usepackage{amssymb}
\usepackage{multirow}

\newcommand{\APSmax}{\ensuremath{\mathrm{A_{PS}^{max}}}}
\newcommand{\APSmin}{\ensuremath{\mathrm{A_{PS}^{min}}}}
\newcommand{\dm}{\ensuremath{\Delta m_d}}
\newcommand{\dt}{\ensuremath{\Delta t}}
\newcommand{\tmin}{\ensuremath{t_{\text{min}}}}

\newcommand{\bmes}{\ensuremath{\mathrm{B}}}

\newcommand{\bplus}{\ensuremath{\mathrm{B}^+}}
\newcommand{\bmin}{\ensuremath{\mathrm{B}^-}}
\newcommand{\bz}{\ensuremath{\mathrm{B}^0}}
\newcommand{\bzbar}{\ensuremath{\overline{\mathrm{B}}{}^{\:0}}}
\newcommand{\dmes}{\ensuremath{\mathrm{D}}}
\newcommand{\dbar}{\ensuremath{\overline{\mathrm{D}}}}
\newcommand{\dstar}{\ensuremath{\mathrm{D}^\ast}}
\newcommand{\dstarm}{\ensuremath{\mathrm{D}^{\ast -}}}
\newcommand{\dz}{\ensuremath{\mathrm{D}^0}}
\newcommand{\dzbar}{\ensuremath{\overline{\mathrm{D}}{}^{\:0}}}
\newcommand{\kmes}{\ensuremath{\mathrm{K}}}
\newcommand{\kmin}{\ensuremath{\mathrm{K}^-}}
\newcommand{\kplus}{\ensuremath{\mathrm{K}^+}}
\newcommand{\ks}{\ensuremath{\mathrm{K}_S^0}}
\newcommand{\kl}{\ensuremath{\mathrm{K}_L^0}}
\newcommand{\kz}{\ensuremath{\mathrm{K}^0}}
\newcommand{\ufours}{\ensuremath{\Upsilon(4S)}}

\newcommand{\electron}{\ensuremath{\mathrm{e}^-}}
\newcommand{\positron}{\ensuremath{\mathrm{e}^+}}
\newcommand{\epem}{\positron \electron}
\newcommand{\BBbar}{\ensuremath{\mathrm{B}\overline{\mathrm{B}}}}

\newcommand{\pb}{\ensuremath{\mathrm{pb}}}
\newcommand{\ps}{\ensuremath{\mathrm{ps}}}

\newcommand{\br}{\ensuremath{\mathcal{B}}}

\newcommand{\ket}[1]{\ensuremath{| #1 \rangle}}
\newcommand{\braket}[2]{\ensuremath{\langle #1 | #2 \rangle}}

\begin{document}

\title		{\boldmath Quantum entanglement at the $\psi(3770)$ and $\Upsilon(4S)$}

\author		{B.D.~Yabsley}
\affiliation	{School of Physics, University of Sydney. NSW 2006, Australia.}

\begin{abstract}
We review results which explicitly depend on the entanglement
of neutral meson pairs produced at the $\psi(3770)$ and 
$\Upsilon(4S)$. Time-dependent CP-violation analyses at the B-factories
use the flavour-singlet final state at the $\Upsilon(4S)$,
but by assuming its quantum-mechanical evolution;
Belle on the other hand has tested the time-dependent flavour correlation
of the \bmes-mesons,
comparing predictions of quantum mechanics, spontaneous disentanglement,
and Pompili-Selleri models.
At the $\psi(3770)$, decay rates are modulated by various combinations
of the charm mixing parameters: this has been exploited by CLEO-c
to provide the first effective constraint on the strong-phase difference
$\delta$. Finally, the goal of a ``model-independent'' $\phi_3$/Dalitz
analysis is now within reach, using \dmes-mesons from the $\psi(3770)$ to
constrain the $\dz\to\ks\pi^+\pi^-$ decay amplitude.
Manifestly entangled events
$\psi(3770) \to (\ks\pi^+\pi^-)_{\dmes} (\ks\pi^+\pi^-)_{\dmes}$,
rather than just ``CP-tagged'' decays, turn out to be crucial.
\end{abstract}

\maketitle

\thispagestyle{fancy}


\section{Introduction}

\begin{figure}
  \includegraphics[width=8cm]{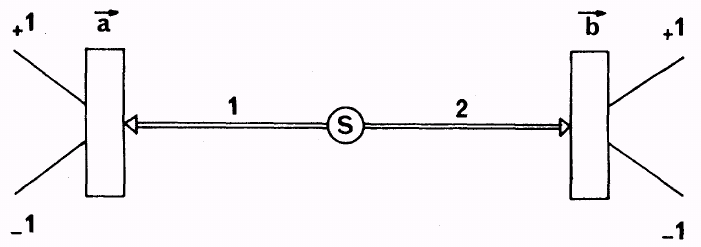}
  \caption{Bohm's version of the EPR ``experiment''. From~\protect\cite{aspect}.}
  \label{fig:EPR-singlet}
\end{figure}

Einstein-Podolsky-Rosen correlations~\cite{epr},
in the form cited by Bohm~\cite{epr-bohm},
are one of the most celebrated features of quantum mechanics.
For a spin-singlet state of photons or particles, 
\begin{equation}
   \frac{1}{\sqrt{2}}
	\left[	\left|	\Uparrow	\right\rangle_1
		\left|	\Downarrow	\right\rangle_2 -
		\left|	\Downarrow	\right\rangle_1 
		\left|	\Uparrow	\right\rangle_2
	\right],
  \label{eq:spin-singlet}
\end{equation}
measurements on particle 1 (2) are indeterminate, 
but once made they fully determine the result of a measurement
on particle 2 (1).
For the arrangement shown in Fig.~\ref{fig:EPR-singlet},
Bell's theorem~\cite{bell} (in the CHSH~\cite{chsh} form)
shows that 
\begin{equation}
  |S|	\equiv \left|
	  E\bigl(\vec{a},\vec{b}\bigr)	
	- E\bigl(\vec{a},\vec{b}^\prime\bigr)
	+ E\bigl(\vec{a}^\prime,\vec{b}\bigr)
	+ E\bigl(\vec{a}^\prime,\vec{b}^\prime\bigr) 
	\right|
	\leq 2
  \label{eq:bell-chsh}
\end{equation}
for any local realistic model; 
quantum mechanics allows for $|S|$ as large as $2\sqrt{2}$
for an optimal choice of settings.\footnote{Here
	$E$ is the correlation of measurements	\\
	$E(\vec{a},\vec{b}) =
		\frac	{R_{++}(\vec{a},\vec{b}) + R_{--}(\vec{a},\vec{b})
			-R_{+-}(\vec{a},\vec{b}) - R_{-+}(\vec{a},\vec{b})}
			{R_{++}(\vec{a},\vec{b}) + R_{--}(\vec{a},\vec{b})
			-R_{+-}(\vec{a},\vec{b}) + R_{-+}(\vec{a},\vec{b})}$.}
Results breaching the bound~(\ref{eq:bell-chsh}) thus rule out local realism, 
even if we subsequently ``get behind'' quantum mechanics to a more complete theory:
the quantum weirdness, or at least this part of it, is a phenomenon of nature.

As shown by Aspect \emph{et al.}~\cite{aspect}, and many times since,
the bound~(\ref{eq:bell-chsh}) is broken for photons produced in cascade decays,
with correlated polarizations:
the data moreover are consistent with quantum mechanics.
(The Aspect result was $S = 2.697 \pm 0.015$, \emph{cf.}\ $S_{QM} = 2.70 \pm 0.05$.)

This behaviour is less well-tested for massive systems,
but is nonetheless routinely used. Decay of a vector meson 
to a pair of neutral pseudoscalars,
\begin{equation}
  \epem \to \ufours \to \frac{1}{\sqrt{2}}\left(
		\ket{\bz}\ket{\bzbar} - \ket{\bzbar}\ket{\bz}
	    \right),
  \label{eq:flavour-singlet}
\end{equation}
produces a \bmes-pair entangled in a \emph{flavour} singlet state:
the flavours of the individual \bmes-mesons are indeterminate,
but at a given time $t$, the pair is always \bz\bzbar.
In the flagship measurements of time-dependent CP violation
at the \bmes-factories,
one \bmes-meson $\bz_{\text{TAG}}$ is reconstructed in a state of definite flavor,
and the other $\bz_{\text{CP}}$ in an eigenstate of CP.
The decay rate is then modulated in the difference of decay times
$\Delta t \equiv t_1 - t_2$,
\begin{multline}
  \Gamma_{CP}(\Delta t) =
	\frac{e^{-|\Delta t|/\tau_{\bz}}}{4\tau_{\bz}}
	\left[ 1 \pm \left\{
	  S_{CP} \sin\left( \Delta m \Delta t \right)	\right.\right.	\\
			 				\left.\left.
	+
	  A_{CP} \cos\left( \Delta m \Delta t \right)
	\right\}
        \right],
  \label{eq:Y4S-cpv}
\end{multline}
with one rate for $\bz_{\text{TAG}}$ ($+$) and another rate for $\bzbar_{\text{TAG}}$ ($-$).
The coefficients $S_{CP}$ and $A_{CP}$ in~(\ref{eq:Y4S-cpv}) are thus CP-violating.

A complementary measurement,
assuming the \bmes-physics but testing the time-dependent flavour oscillation
due to the entanglement~(\ref{eq:flavour-singlet}),
has been performed by Belle (Section~\ref{section-belle}).
In the formally equivalent, but experimentally different setting of $\psi(3770)$ decays,
decay rates of \dmes-meson pairs are modulated by the charm mixing parameters 
$(x,y)$ and the strong phase difference $\delta$: CLEO-c has exploited this
to derive a constraint on the latter quantity (Section~\ref{section-cleo-c}).
The final state $\psi(3770) \to (\ks\pi^+\pi^-)_{\dmes} (\ks\pi^+\pi^-)_{\dmes}$,
where the entanglement of the \dmes-mesons is clearly manifest,
also turns out to be crucial for the ``model-independent'' measurement 
of the unitarity angle $\phi_3$ using the Dalitz analysis method
at the B-factories (Section~\ref{section-phi3}).


\section{\boldmath \ufours: EPR correlations at Belle}
\label{section-belle}

\begin{figure}
  \includegraphics[width=8.0cm]{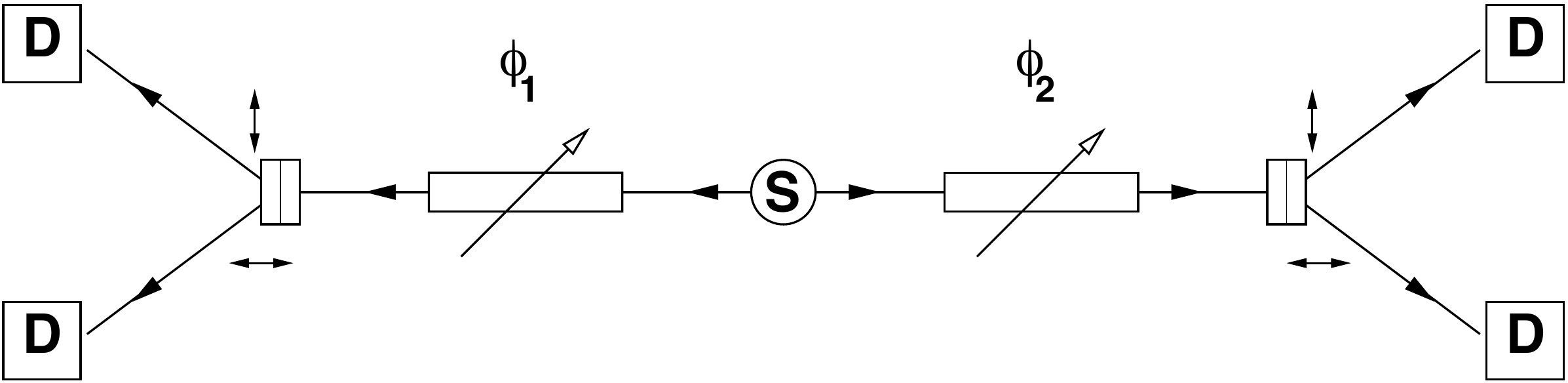}
  \caption{An optical analogue of the $\ufours\to\BBbar$
	 EPR correlation analysis.}
  \label{fig:optical-analogue}
\end{figure}

In the \emph{quasi-spin} analogy initially expounded
for \kmes-mesons~\cite{quasispin},
a $\ket{\bz}$ corresponds to a spin $\ket{\!\Uparrow}_z$ particle
or a photon with vertical polarization (V);
a $\ket{\bzbar}$ corresponds to a spin $\ket{\!\Downarrow}_z$ particle
or a horizontally polarized photon (H).
While optical measurements can be made on arbitrary axes
$\alpha\text{$\ket{\!\Uparrow}$} + \beta\text{$\ket{\!\Downarrow}$}$,
only the $\ket{\!\Uparrow}$ and $\ket{\!\Downarrow}$
measurements are practical for \bmes-mesons.
However, over time $t$ the state $\ket{\bz}$ evolves to
\begin{equation}
  \frac{1}{2}\left[
	\{1+\cos(\dm t)\}\ket{\bz} + \{1-\cos(\dm t)\}\ket{\bzbar}
  \right],
  \label{eq:b-time-evolution}
\end{equation}
making other ``measurement axes'' accessible. Thus for \bmes-pairs
produced in the flavour singlet state~(\ref{eq:flavour-singlet}),
the \emph{decay time difference} $\dm\dt$ plays the role of the difference
$\Delta\phi$ between polarimeter orientations in an optical experiment.
An optical analogue of the situation at the \bmes-factories is shown
in Fig.~\ref{fig:optical-analogue}: 
the polarimeters are fixed such that they always perform a V/H measurement
on the photons, but phase rotations $\phi_1$ and $\phi_2$ are inserted 
between the production point of the entangled photons and their measurement.

This setup would seem to allow a full Bell inequality test, 
but in fact the rotations $\phi_{1,2}$ are imposed by nature---in the decay
times of the particles---rather than being subject to the experimenter's choice.
Thus a true Bell test cannot be performed.

\subsection{The Green Baize Table Conspiracy Model}

While there are general arguments for the impossibility
of such a test~\cite{bertlmann},\footnote{It turns out that even in the case
	of active flavor measurement (rather than letting the mesons decay),
	a Bell inequality test cannot be performed,
	because the decay of the B-mesons is too rapid compared to their
	flavour oscillation,
	\emph{i.e.}\ $x_d = \Delta m_d / \Gamma_d \simeq 0.77$ is too small.}
it is more entertaining to consider counter-examples. 
A breathtakingly \emph{unrealistic} local-realistic model,
indistinguishable in its predictions from quantum mechanics,
was devised for this purpose by Bramon, Escribano, and Garbarino~\cite{bramon}
(following Kasday~\cite{kasday}): we expound it here in a form
inspired by one of the myths of our age.

\begin{figure}
  \includegraphics[width=6cm]{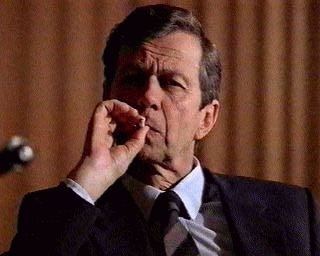}
  \caption{According to one local-hidden-variable model,
	the Cigarette-Smoking Man takes a close interest in $\ufours\to\BBbar$ decays.}
  \label{fig:cancerman}
\end{figure}

Somewhere, there is a wood-panelled room with a green baize table,
where powerful men meet together, smoke, and make conspiracy
(Fig.~\ref{fig:cancerman}).
They determine world events in detail, including decays $\ufours\to\BBbar$.
At the time of each decay, $t=0$, four hidden variables are set:
mesons 1 and 2 are each given a piece of paper (as it were)
bearing the quantities $(t_1,f_1)$ and $(t_2,f_2)$.
These act locally: meson $i$ decays at a time $t = t_i$,
into final state $f = f_i$, according to the values
written on its piece of paper.
Now as part of the conspiracy, in order to deceive the world,
$(t_1,f_1,t_2,f_2)$ are chosen randomly according to quantum mechanical rules.

The phenomena in this model are indistinguishable from those in quantum mechanics,
even though the individual particles have a definite state at all times:
the model is local, and (if only in this sense) realistic.\footnote{Optical
	experiments do not share this vulnerability.
	For example, in the modification of the Aspect experiment
	reported by Weihs \emph{et al.}~\protect\cite{weihs},
	the effective polarimeter orientation
	is changed while the photons are in flight,
	according to randomly generated numbers.
	In this case, no conspiracy can fix the results to conform to
	a particular pattern in $\Delta \phi \equiv \dm\dt$.}
Another explicit counter-example has since been constructed by Santos~\cite{santos}.

\subsection{QM versus specific LR models}

\begin{figure}
  \includegraphics[width=5.2cm]{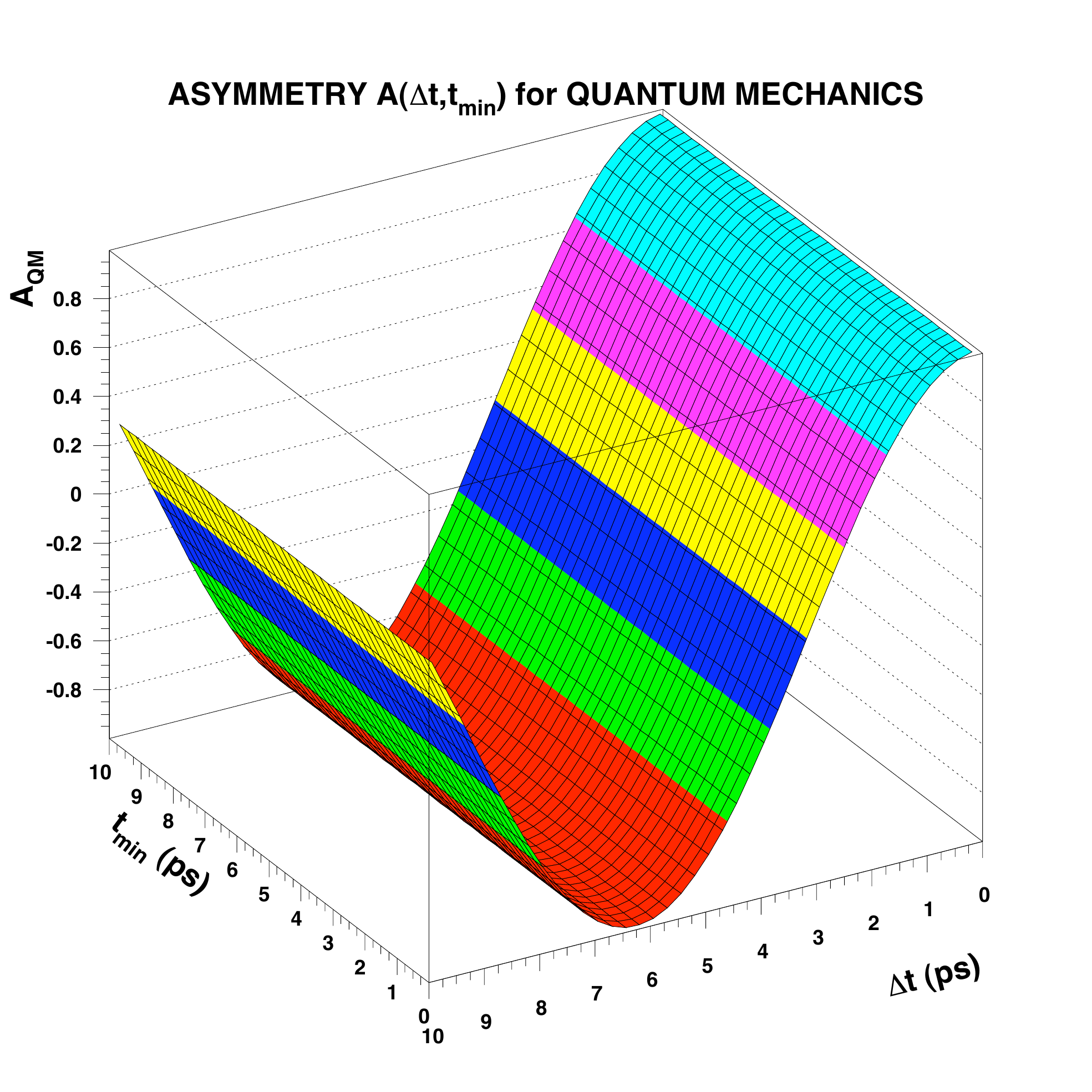}	\\[0.0ex]
  \includegraphics[width=5.2cm]{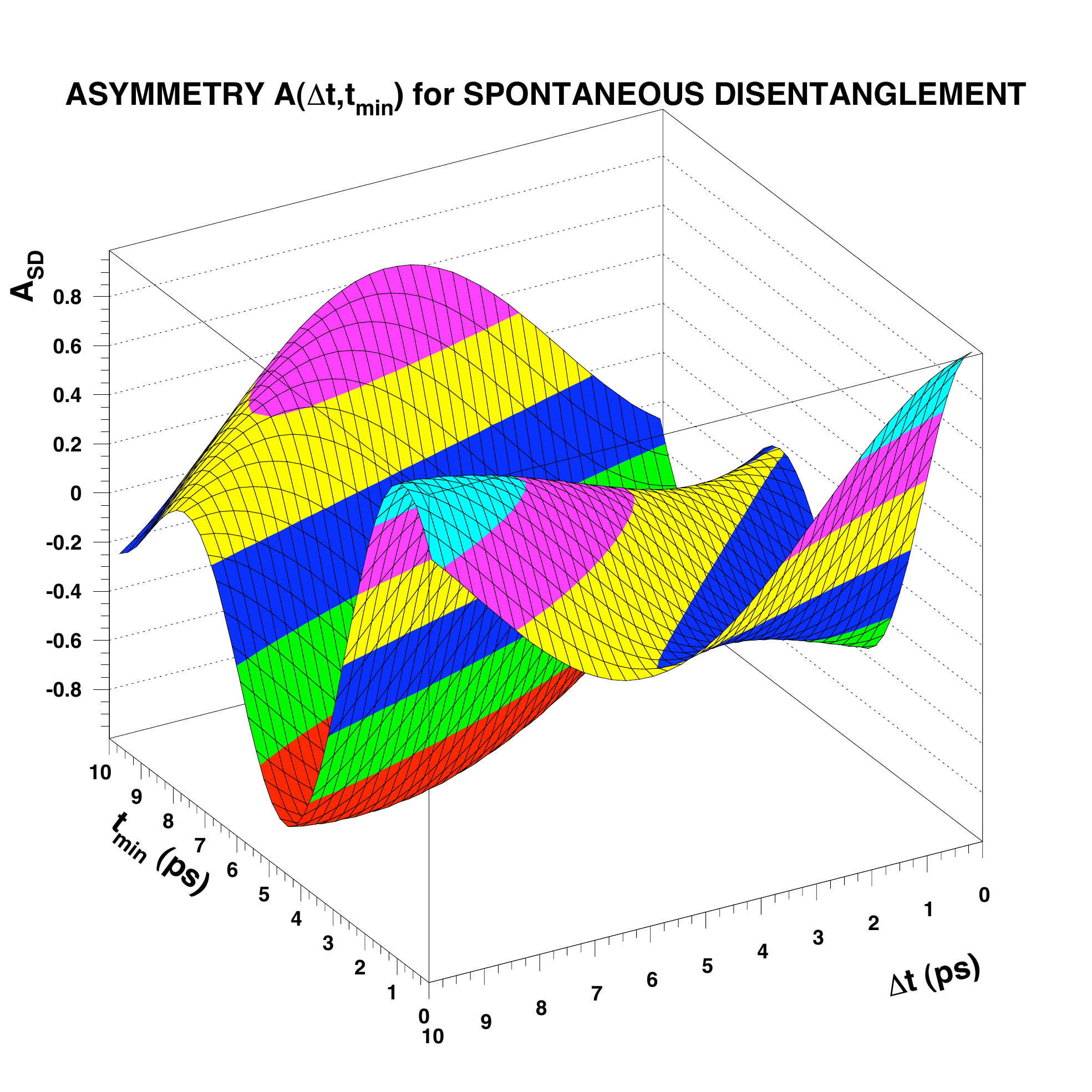}	\\[0.0ex]
  \includegraphics[width=5.2cm]{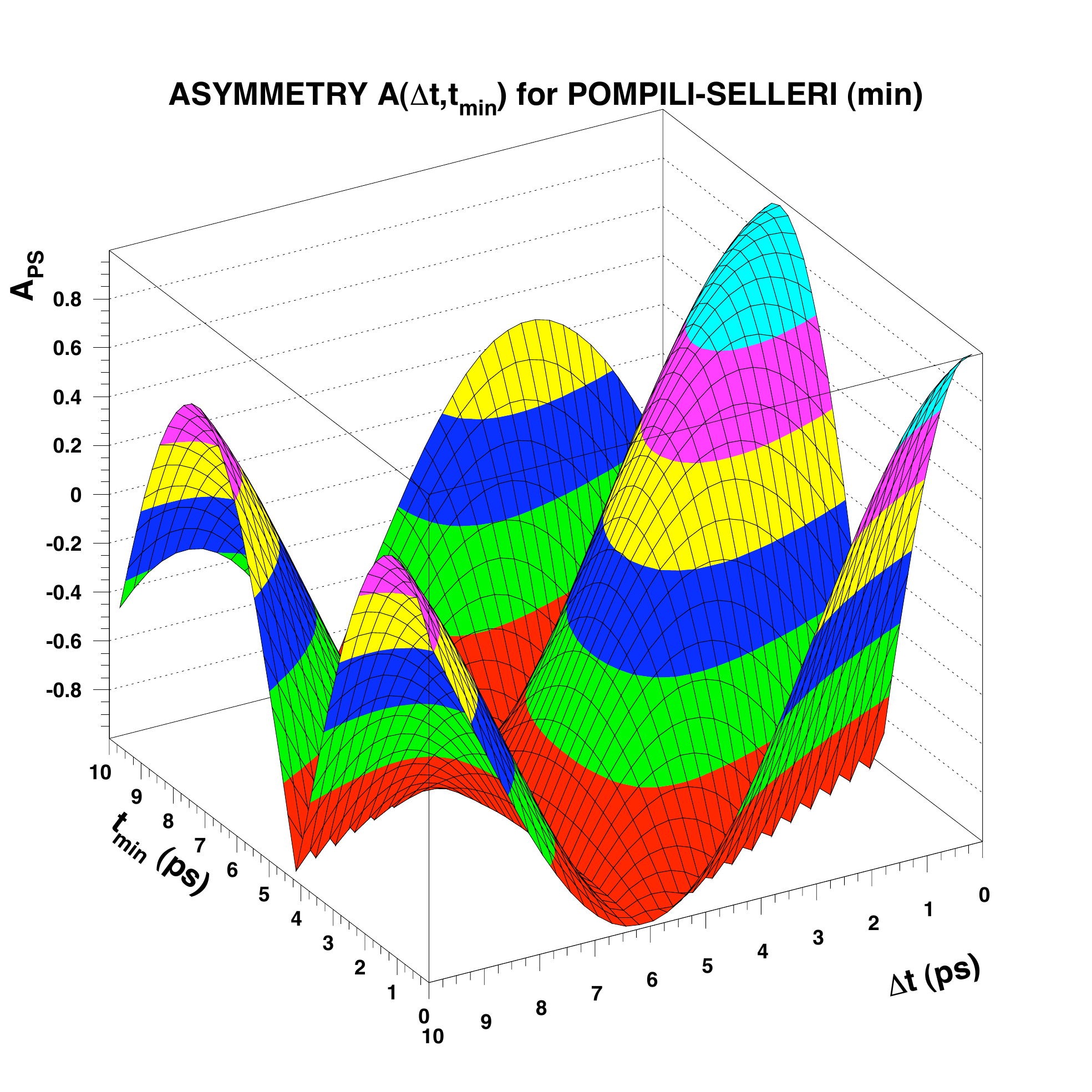}\\[0.0ex]
  \includegraphics[width=5.2cm]{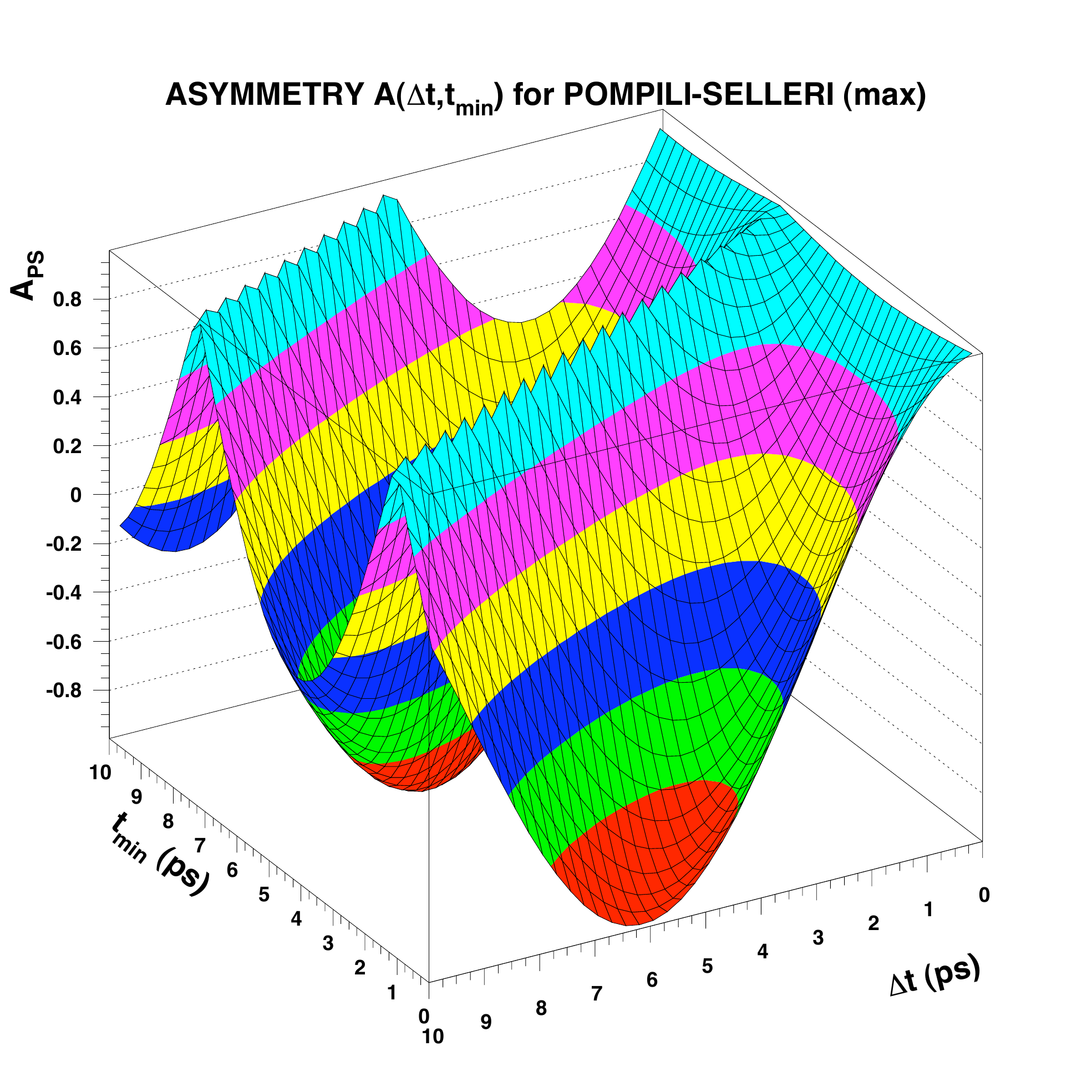}
  \caption{For ideal measurement at $\ufours\to\BBbar$:
	decay rate asymmetry $A = (R_{OF} - R_{SF}) / (R_{OF} + R_{SF})$
	as a function of $(\dt,\tmin)$,
	(top) for quantum mechanics;
	(second) for spontaneous disentanglement;
	and for the Pompili-Selleri class of models, showing
	(third) minimum, and (bottom) maximum values.}
  \label{fig:asymmetry-dt-tmin}
\end{figure}

The best that can be done is thus to test the data in detail against
quantum mechanics, and the predictions of specific local realistic models. 
Reasonable, non-conspiratorial models can then potentially be ruled out.
The experimental quantity of choice is the time-dependent asymmetry
of decay rates to opposite-flavour (OF) and same-flavor (SF) final states:
\begin{align}
  A(t_1,t_2)	& \equiv \frac	{R_{OF}(t_1,t_2) - R_{SF}(t_1,t_2)}
				{R_{OF}(t_1,t_2) + R_{SF}(t_1,t_2)}	
  \label{eq:asymmetry}							\\
		& = \cos(\dm\dt)
  \label{eq:asymmetry-qm}
\end{align}
for quantum mechanics. The dependence on $\dt = t_1 - t_2$ alone,
a consequence of the entanglement of the state, is distinctive.
It's therefore useful to consider asymmetries as functions not of $(t_1,t_2)$,
but of \dt\ and $\tmin \equiv \min(t_1,t_2)$. Several possibilities
for $A(\dt,\tmin)$ are shown in	Fig.~\ref{fig:asymmetry-dt-tmin}.
As a limiting case, under spontaneous disentanglement (SD) to a $\bz\bzbar$ pair
(with definite flavour) immediately after \ufours\ decay,
the two mesons undergo independent flavor oscillations, with
\begin{equation}
  A_{SD} = \cos(\dm t_1)\cos(\dm t_2),
  \label{eq:asymmetry-sd}
\end{equation}
taking the complicated form in the figure when plotted on $(\dt,\tmin)$.
More seriously, one can consider the class of models obeying the assumptions
of Pompili and Selleri~\cite{pompili-selleri}:
QM-like states for the individual mesons with stable mass;
100\% flavour correlations, to reproduce the QM behaviour as closely as possible;
and the constraint that QM predictions for uncorrelated \bmes-mesons should
also be preserved. Asymmetries must then lie between the bounds
\begin{align}
\APSmin	& =	1-\min(2+\Psi,2-\Psi),\;\text{where}		    \nonumber	\\
\Psi	& =	\left\{1+\cos(\dm \Delta t)\right\}\cos(\dm \tmin)  \nonumber	\\
	& \phantom{=}\,
		\phantom{1}\, - \sin(\dm \Delta t)\sin(\dm \tmin)
  \label{eq:asymmetry-psmin}							\\
\intertext{and}
\APSmax	& = 1- |\left\{1-\cos(\dm \Delta t)\right\}\cos(\dm \tmin)  \nonumber	\\
	& \phantom{=}\,
		\phantom{1}\, + \sin(\dm \Delta t)\sin(\dm \tmin)|
  \label{eq:asymmetry-psmax}
\end{align}
shown in the figure.

In the current state of the art, the discrimination power
shown in Fig.~\ref{fig:asymmetry-dt-tmin} is not fully realised,
as only \dt\ is measured:
the expressions~(\ref{eq:asymmetry-sd})--(\ref{eq:asymmetry-psmax})
must in effect be integrated over \tmin.
It turns out that it is still possible to exclude the local-realistic models shown.

\subsection{The 2007 Belle result}

\begin{figure*}
  \includegraphics[width=16cm]{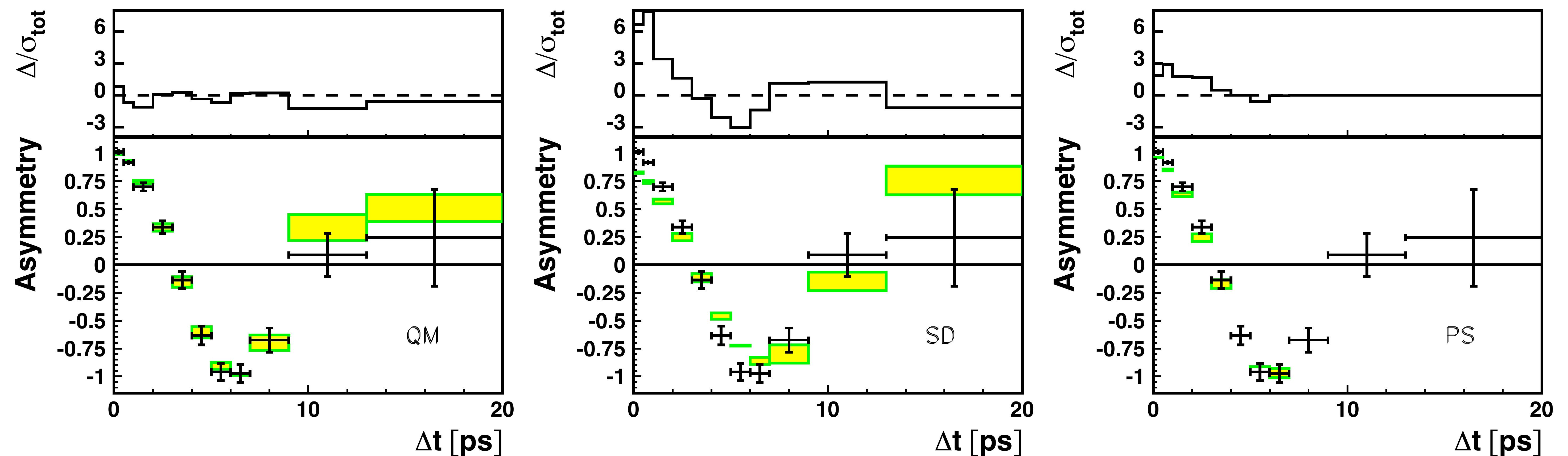}
  \caption{From~\protect\cite{belle-EPR}:
	Asymmetries as a function of \dt\ in Belle data,
	compared to (left) quantum mechanics, (middle) spontaneous disentanglement, 
	and (right) Pompili-Selleri models. 
	The lower plots show time-dependent flavour asymmetry (crosses) 
	and the results of weighted least-squares fits to each model
	(rectangles, showing $\pm 1 \sigma$ errors on \dm).	
	Upper plots show differences $\Delta \equiv A_{\text{data}}-A_{\text{model}}$
	in each bin, divided by the total experimental error $\sigma_{\text{tot}}$.
	Bins where $\APSmin < A_{\text{data}} < \APSmax$
	have been assigned a null deviation.}
  \label{fig:Belle-EPR-fit}
\end{figure*}

The Belle analysis~\cite{belle-EPR} is based on
a $152\times 10^6$ \BBbar\ data sample, and uses techniques established
for time-dependent CP-violation analyses:
one \bmes-meson is reconstructed in a flavour-tagging mode
$\bz\to \dstarm\ell^+\nu$,
while the other \bmes-flavour is tagged using a lepton;
a consistency check with other flavour-tagging information is imposed
to maintain purity. A sample of 8565 such events is found:
6718 opposite-flavour, and 1847 same-flavour pairs,
divided into 11 bins in \dt.

Backgrounds are subtracted from the OF and SF samples separately, in \dt\ bins: 
fake \dstar, using sidebands ($126\pm 6$ OF, $54\pm 4$ SF);
bad $\dstar-\ell$ combinations, also from data ($78\pm 9$ OF, $236\pm 15$ SF); and
$\bplus\to\dbar^{**0}\ell\nu$ events, estimated from Monte Carlo
and the $\cos_{\bmes,\dstar\ell}$ distribution in data
(254 OF and 1.5 SF [both $\pm 6\%$]).
This process is crucial, as the backgrounds produce
a time-structured difference in the asymmetry.
After a $(1.5\pm 0.5)\%$ correction for mistagging,
deconvolution is performed (using DSVD, on the SF and OF samples separately)
to remove decay-vertex-resolution, efficiency, and other remaining effects.
(Potential bias against any of the models is explicitly studied
and subtracted, with a systematic error assigned.)
Finally, a simple exponential fit to the resulting histogram is performed,
to extract the \bmes-lifetime as a check:
$\tau_{\bz} = (1.532\pm 0.017)\,\ps$ is found,
\emph{cf.}\ the $(1.530 \pm 0.009)\,\ps$ world average~\cite{pdg2006}.

A comparison between the data, and fits to the predictions of each model,
is shown in Fig.~\ref{fig:Belle-EPR-fit}.
In each case, the parameter \dm\ is floated,
subject to the constraint of the world average value,
with Belle and BaBar contributions excluded:
$(0.496\pm 0.014)\,\ps^{-1}$~\cite{hfag2006}.\footnote{The Belle
	and BaBar results are from fits assuming QM time evolution,
	and so cannot be used. They otherwise dominate the average, 
	improving its precision by a factor of three.}
Quantum mechanical predictions fit the data well ($\chi^2/n_{dof} = 5/11$),
while spontaneous disentaglement is disfavoured at $13\sigma$ ($\chi^2 = 174$).
A fit to $(1-\zeta_{\bz\bzbar}) A_{QM}(\dt) + \zeta_{\bz\bzbar} A_{SD}(\dt)$
finds an ``SD fraction'' of $\zeta_{\bz\bzbar} = 0.029 \pm 0.057$,
consistent with zero.\footnote{In a decoherence model,
	this is equivalent to multiplying the interference term
	in the \bz-\bzbar\ basis by a factor
	$(1-\zeta_{\bz\bzbar})$~\protect\cite{bertlmann-decoherence}.}
The entire class of local realistic models satisfying the minimal
Pompili-Selleri assumptions is disfavoured at $5.1\sigma$ ($\chi^2 = 31$).


\section{\boldmath $\psi(3770)$: $(x,y,\delta)$ and rates at CLEO-c}
\label{section-cleo-c}

Formally, the situation at the $\psi(3770)$ is the same as that
at the \ufours: the decay
$\epem \to \psi(3770) \to \frac{1}{\sqrt{2}}
				\left(
				  \ket{\dz}\ket{\dzbar} - \ket{\dzbar}\ket{\dz}
				\right)$
leads to an entangled final state equivalent to~(\ref{eq:flavour-singlet}).
However there are differences in practice:
mixing is a percent-level effect in \dmes-decay amplitudes,
and CP violation is suppressed orders of magnitude further. 
So while the principal use of the state~(\ref{eq:flavour-singlet})
at the \ufours\ is study of CP violation,
an obvious use of \dmes-meson pairs produced at the $\psi(3770)$
is CP tagging. For example, decays to two CP-even (or two CP-odd)
eigenstates don't occur.

If we consider decays
$\psi(3770) \to (\kmin\pi^+)_{\dmes} (\kmin\pi^+)_{\dmes}$,
however, the situation is not so straightforward.
Relative to production from a pair of \dz\ mesons, 
the rate is suppressed by the \emph{mixing rate} $R_M = \frac{1}{2}(x^2+y^2)$;
by contrast, the rate for uncorrelated $\dmes\dbar$ decays to this final state
is suppressed by only the ``wrong-sign'' rate $R_{WS}$,
\emph{i.e.}\ it is forty times larger.
There are thus nontrivial effects due to the coherence of the state
produced at the $\psi(3770)$. 
Currently the most systematic treatment is~\cite{asner-sun}, 
following earlier work by~\cite{gronau-grossman-rosner} and others.

\subsection{\boldmath $\dz\to \kz_{S,L} \pi^0$}

A simple example is the study of $\dz \to \kl \pi^0$
by CLEO-c~\cite{cleo-klpi0}, where the $\dz \to \kl \pi^0$
decay is recovered using the distribution of missing-mass-squared
$M_{\text{miss}}^2$ in events tagged by a fully-reconstructed \dzbar\ decay.
In pratice there are three distinct samples,
corresponding to the three tagging modes
$\dzbar\to\kplus\pi^-$, $\kplus\pi^-\pi^0$, and $\kplus\pi^-\pi^-\pi^+$.
A $\dz \to \kl \pi^0$ branching fraction calculation using tagging mode $f$
in fact determines
\[
  \br_{\kl\pi^0}
	\left(
		1 + \frac{2 r_f \cos \delta_f + y}{1 + R_{WS,f}}
	\right),
\]
where the amplitude ratio $r_f$ and strong phase difference $\delta_f$
($r_f e^{-i\delta_f} \equiv \braket{f}{\dzbar}/\braket{f}{\dz}$)
and the wrong-sign rate $R_{WS,f}$ are mode-dependent. 

The method is to use the equivalent measurement for $\dz\to\ks\pi^0$,
an an untagged measurement of the branching fraction for that mode,
to determine the product $C_f = (2 r_f \cos \delta_f + y)/(1 + R_{WS,f})$.
(A tagged analysis in this mode finds an effective branching fraction
$\br_{\ks\pi^0}(1 - C_f)$.)
The true $\dz\to\kz\pi^0$ branching fraction can then be measured for
each tagged sample: averaging over them, CLEO-c find
\begin{equation}
  \br_{\kl\pi^0} = (0.998\pm 0.049\pm 0.030\pm 0.038)\%,
  \label{eq:klpi0-branching}
\end{equation}
and an asymmetry
\begin{equation}
  \frac	{\br_{\ks\pi^0}-\br_{\kl\pi^0}}
	{\br_{\ks\pi^0}+\br_{\kl\pi^0}}	= 0.108 \pm 0.025 \pm 0.024,
  \label{eq:klpi0-asymmetry}
\end{equation}
consistent with the value $2\tan\theta_C = 0.109 \pm 0.001$
expected if symmetry under the U-spin subgroup of $\mathrm{SU(3)}$
is imposed.

\subsection{\boldmath Charm mixing and $\delta$}

\begin{table}
  \caption{From~\protect\cite{cleo-delta-kpi}.
	Effective branching fractions 
	(upper section) for \dz\ decay modes $i$, divided by $\br_i$, and
	(lower section) for \dz\dzbar\ decay to modes $\{i,j\}$,
			divided by $\br_i \br_j$,
	in $\psi(3770) \to \frac{1}{\sqrt{2}}\left(
			\ket{\dz}\ket{\dzbar} - \ket{\dzbar}\ket{\dz}
	    	\right)$ data.
	$S_+$ ($S_-$) denotes a CP-even (CP-odd) eigenstate,
	and \electron\ a semileptonic final state $X^+ \electron \bar\nu_e$.
	Quantities are shown to leading order in the mixing parameters $(x,y)$
	and the wrong-sign rate $R_{WS}$.
	$R_M = \frac{1}{2}(x^2 + y^2)$ is the mixing rate.}
  \label{tab:cleo-c-rates}
\begin{tabular}{ccc}
\hline\hline
Mode & Correlated & Uncorrelated \\
\hline
$K^-\pi^+$ &
        $1+R_{\rm WS}$ &
        $1+R_{\rm WS}$ \\
$S_+$ & $2$ & $2$ \\
$S_-$ & $2$ & $2$ \\
\hline
$K^-\pi^+, K^-\pi^+$ &
        $R_{\rm M}$ &
        $R_{\rm WS}$ \\
$K^-\pi^+, K^+\pi^-$ &
	$[(1+R_{\rm WS})^2$ &
        $1+R_{\rm WS}^2$ \\
  &	$-4r\cos\delta(r\cos\delta+y)]$ & \\
$K^-\pi^+, S_+$ &
        $1+ R_{\rm WS}+2r\cos\delta+y$ &
        $1+R_{\rm WS}$ \\
$K^-\pi^+, S_-$ &
        $1+R_{\rm WS}- 2r\cos\delta-y$ &
        $1+R_{\rm WS}$ \\
$K^-\pi^+, e^-$ &
        $1-ry\cos\delta-rx\sin\delta$ &
        $1$ \\
$S_+, S_+$ & 0 & $1$ \\
$S_-, S_-$ & 0 & $1$ \\
$S_+, S_-$ &
        $4$ &
        $2$ \\
$S_+, e^-$ &
        $1+y$ &
        $1$ \\
$S_-, e^-$ &
        $1-y$ &
        $1$\\
\hline\hline
\end{tabular}
\end{table}

In the general case, effective branching fractions depend on 
the mixing parameters $(x,y)$, and (for the $\kmin\pi^+$ final state)
the strong phase difference $\delta \equiv \delta_{\kmes\pi}$.
Correction factors are summarized in Table~\ref{tab:cleo-c-rates}.
Whereas in the $\dz\to\kl\pi^0$ analysis the corrections
were a complication that needed to be taken into account, 
the CLEO-c analysis reported in~\cite{cleo-delta-kpi} uses
a suite of measurements,
and their varying dependence on $(x,y)$ and $\delta$,
to constrain the mixing parameters, and in particular $\delta$.

In the first case a least-squares fit is performed to 
the yields in eight hadronic final states
($\kmin\pi^+$, $\kplus\pi-$, $\kplus\kmin$, $\pi^+\pi^-$,
 $\ks\pi^0\pi^0$, $\ks\pi^0$, $\ks\eta$, and $\ks\omega$),
and 43 ``double-tagged'' final states
(24 fully-reconstructed, 14 including a semileptonic decay,
and 5 including $\kl\pi^0$),
together with external results on seven branching fractions
(CP eigenstates, and $\kmin\pi^+$, with correlations taken 
into account). 
The result finds the lifetime difference parameter $y$ with
large uncertainty, and as a result the combination $x\sin\delta$
is unconstrained. 

\begin{figure}
  \includegraphics[width=7.0cm]{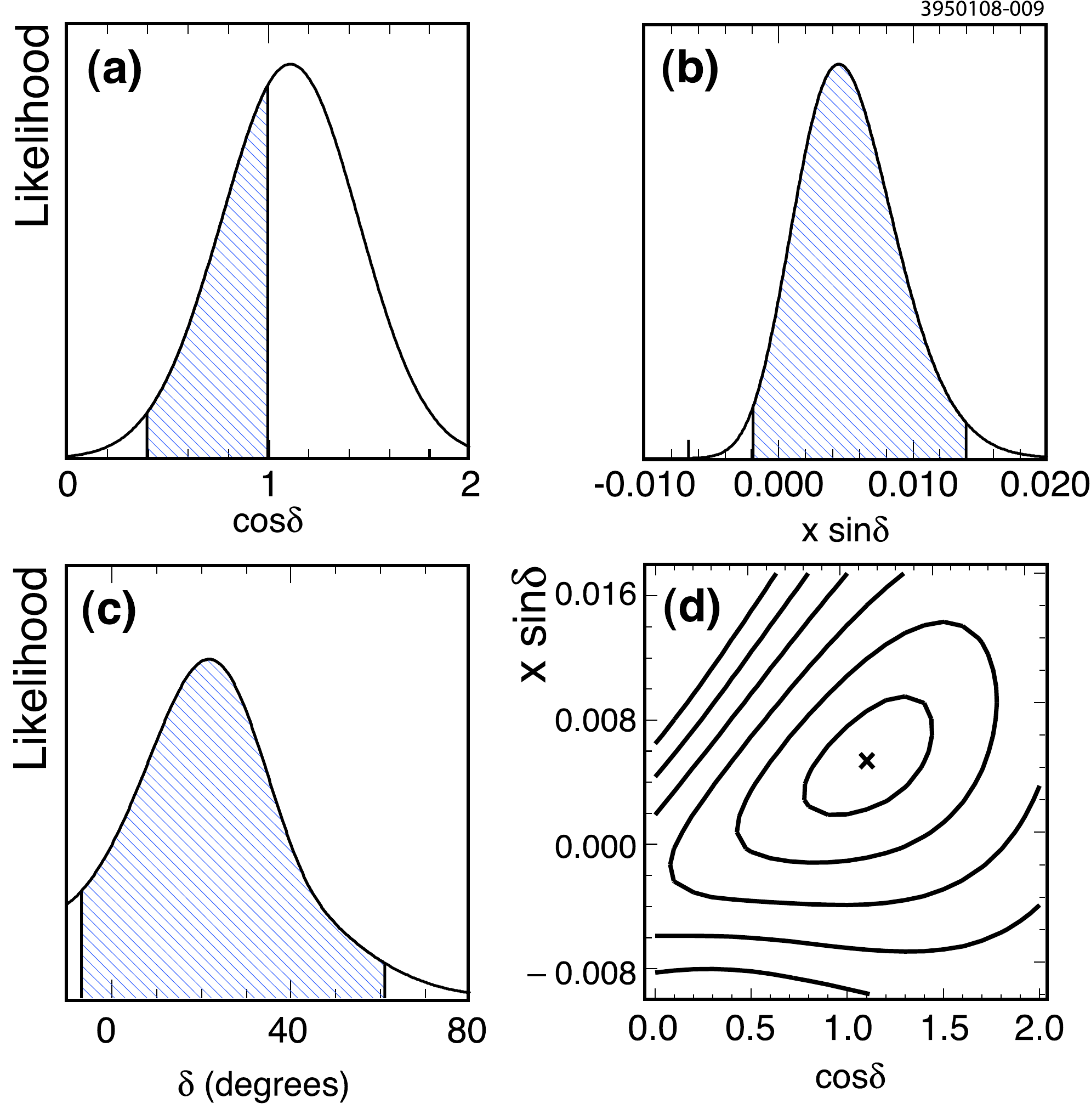}
  \caption{From~\protect\cite{cleo-delta-kpi}: 
	Constraints on the strong phase difference $\delta$, after
	combining results from CLEO-c's entangled source of D-mesons with
	external branching fraction and mixing parameter measurements.
	Likelihood curves (including statistical and systematic uncertainties)
	are shown for (a) $\cos\delta$, (b) $x\sin\delta$, and (c) $\delta$;
	(d) shows contours in units of $\sqrt{\Delta\chi^2}$
	on $(\cos\delta,x\sin\delta)$.}
  \label{fig:cleoc-delta-constraints}
\end{figure}

An extended fit, including measurements of mixing-related quantities
($y,\, x,\, r^2,\, y',\, (x')^2$) by other experiments,
is therefore performed: results are shown in 
Fig.~\ref{fig:cleoc-delta-constraints}.
The fit finds
\begin{align}
  \phantom{x}\cos\delta	& = 1.10 \pm 0.35 \pm 0.07	\nonumber	\\
	x\sin\delta	& = (4.4^{+2.7}_{-1.8} \pm 2.9) \times 10^{-3},
  \label{eq:cleo-c-cosdelta}
\end{align}
and after minimising on the physical surface $(\cos\delta,\sin\delta)$,
\begin{align}
 \delta	& = (22^{+11}_{-12} {}^{+9}_{-11})^\circ,\;\text{with}\nonumber	\\
 \delta & \in [-7^\circ,+61^\circ]\;\text{at 95\% confidence},
 \label{eq:cleo-c-delta}
\end{align}
the first effective constraint on this quantity.
Together with the external mixing measurements
(without which the analysis is not possible),
the precision on $\cos\delta$ is driven by
the yields for the eight hadronic final states,
and double-tag yields $\{\kmes\pi,S_{\pm}\}$ including
CP-eigenstates $S_{\pm}$.
These results are for a $281\,\pb^{-1}$ sample:
better precision will be possible with the final CLEO-c dataset.


\section{\boldmath $\phi_3$/Dalitz: $\psi(3770)$ rescues the \ufours}
\label{section-phi3}

\begin{figure}
  \includegraphics[width=6.5cm]{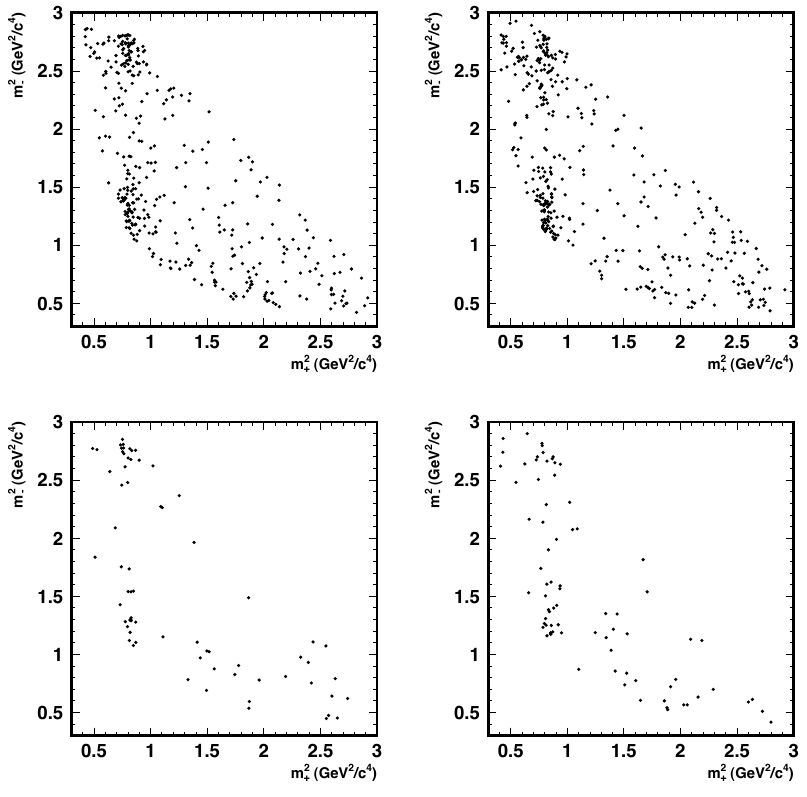}
  \caption{From~\protect\cite{belle-phi3-update08}:
	(preliminary) updated Belle results.
	Dalitz distributions for $\dzbar\to\ks\pi^+\pi^-$ decays
	from (top) $\bmes^\pm\to\dmes\kmes^\pm$ 
	and (bottom) $\bmes^\pm\to\dstar\kmes^\pm$, 
	for (left) \bplus\ and (right) \bmin\ samples.}
  \label{fig:belle-phi3}
  \vspace*{2.5ex}
  \includegraphics[width=6.0cm]{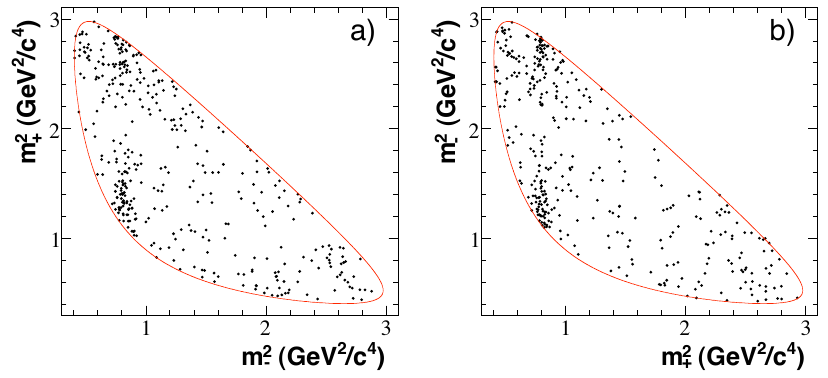} \\[0.5ex]
  \includegraphics[width=6.0cm]{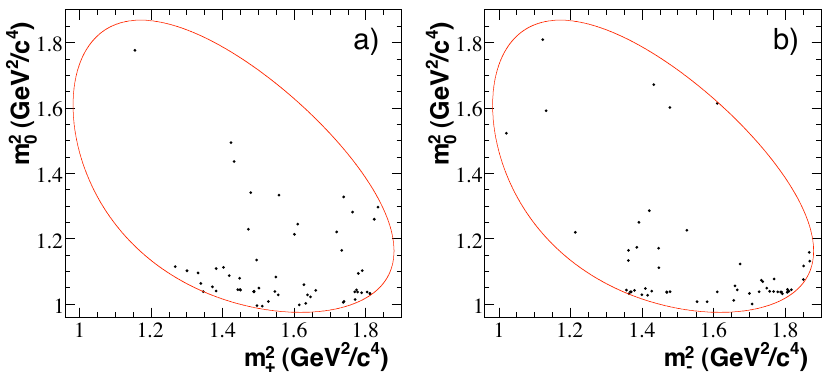} 
  \caption{From~\protect\cite{babar-phi3-update08}:
	a selection of (published) BaBar results.
	Dalitz distributions of 
	(Upper) $\dzbar\to\ks\pi^+\pi^-$
		for (a) $\bmin\to\dmes\kmin$
		and (b) $\bplus\to\dmes\kplus$; and
	(lower) $\dzbar\to\ks\kplus\kmin$ 
		for (a) $\bmin\to\dmes\kmin$
		and (b) $\bplus\to\dmes\kplus$.}
  \label{fig:babar-phi3}
  \vspace*{2.5ex}
  \includegraphics[width=6.0cm]{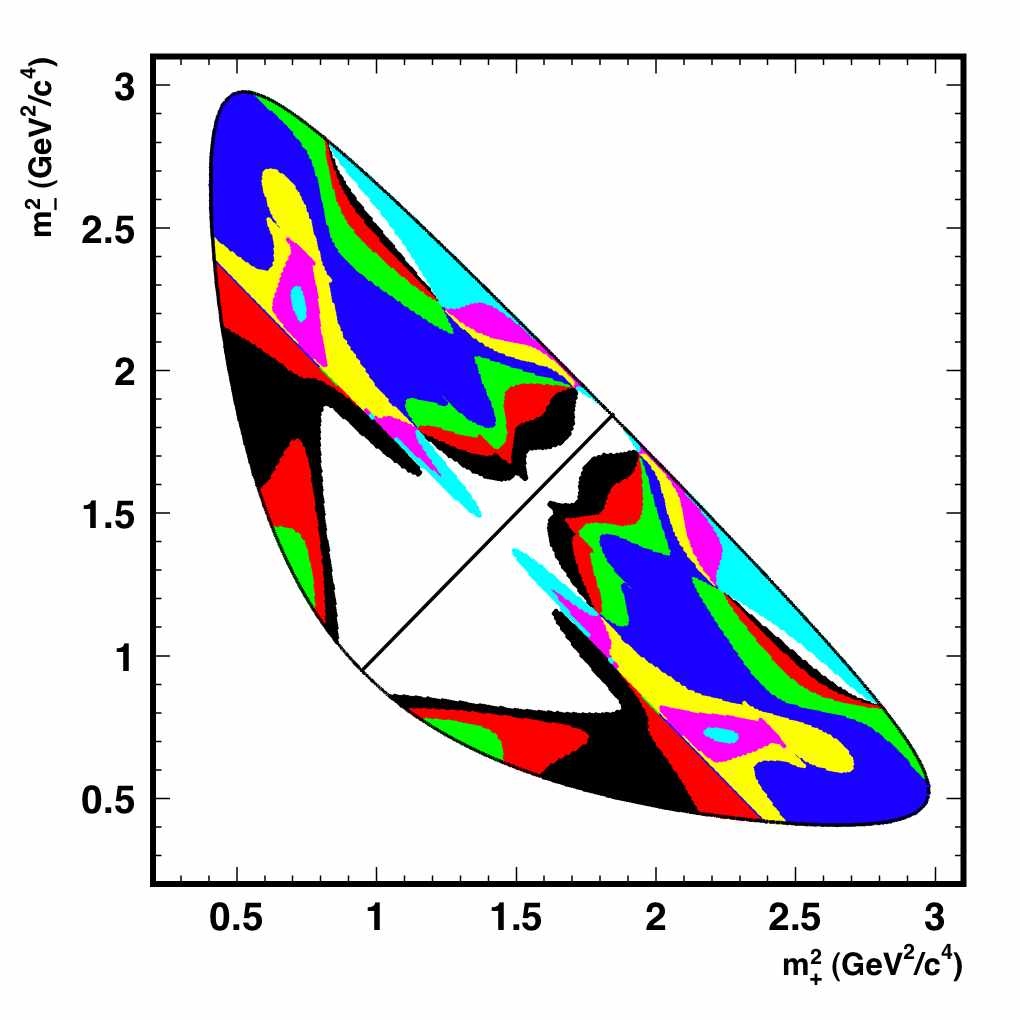}
  \caption{From~\protect\cite{bondar-poluektov-2008}:
	Uniform binning in $\left. \Delta \delta_D \right|_{model}$
	on the $\dzbar\to\ks\pi^+\pi^-$ Dalitz plot; see the text.}
  \label{fig:phi3-delta-binning}
\end{figure}

Dalitz analyses of $\bmes^\pm \to \dmes\kmes^\pm$ and $\dstar\kmes^\pm$,
with $\dmes \to \ks\pi^+\pi^-$ and $\ks\kplus\kmin$,
are currently the most sensitive probe of the unitarity angle $\phi_3$
(also known, in the least interesting of the disagreements between
the B-factories, as $\gamma$).
The state of play in these important analyses was shown in 
Anton Poluektov's talk on Monday~\cite{fpcp08-anton};
Belle's preliminary update~\cite{belle-phi3-update08},
shown in Fig.~\ref{fig:belle-phi3}, finds
$\phi_3 = (76^{+12}_{-13} \pm 4 \pm 9)^\circ$;
BaBar's new publication~\cite{babar-phi3-update08},
shown in Fig.~\ref{fig:babar-phi3}, finds
$\phi_3 = (76 \pm 22 \pm 5 \pm 5)^\circ$.
In both cases, the model error (shown last) is already uncomfortably large.

The future of this measurement is
the so-called model-independent approach~\cite{bondar-poluektov-2006};
the feasibility study for the method has recently been updated~\cite{bondar-poluektov-2008}.
Rather than relying directly on (say) an isobar model to determine
the $\dzbar\to\ks\pi^+\pi^-$ decay amplitude, the phase difference parameters
\begin{align}
  c & = \cos(\delta_D(m_+^2,m_-^2) - \delta_D(m_-^2,m_+^2))	\nonumber \\
  s & = \sin(\delta_D(m_+^2,m_-^2) - \delta_D(m_-^2,m_+^2))
  \label{eq:c-s}
\end{align}
are extracted from $\psi(3770)$ data, exploiting the correlations in the final state
to measure the Dalitz plot, not for \dz\ or \dzbar, but for CP-tagged \dmes-mesons.
The challenge is to cope effectively with the limited data samples available
(or foreseen).

\begin{figure}
  \includegraphics[width=7.5cm]{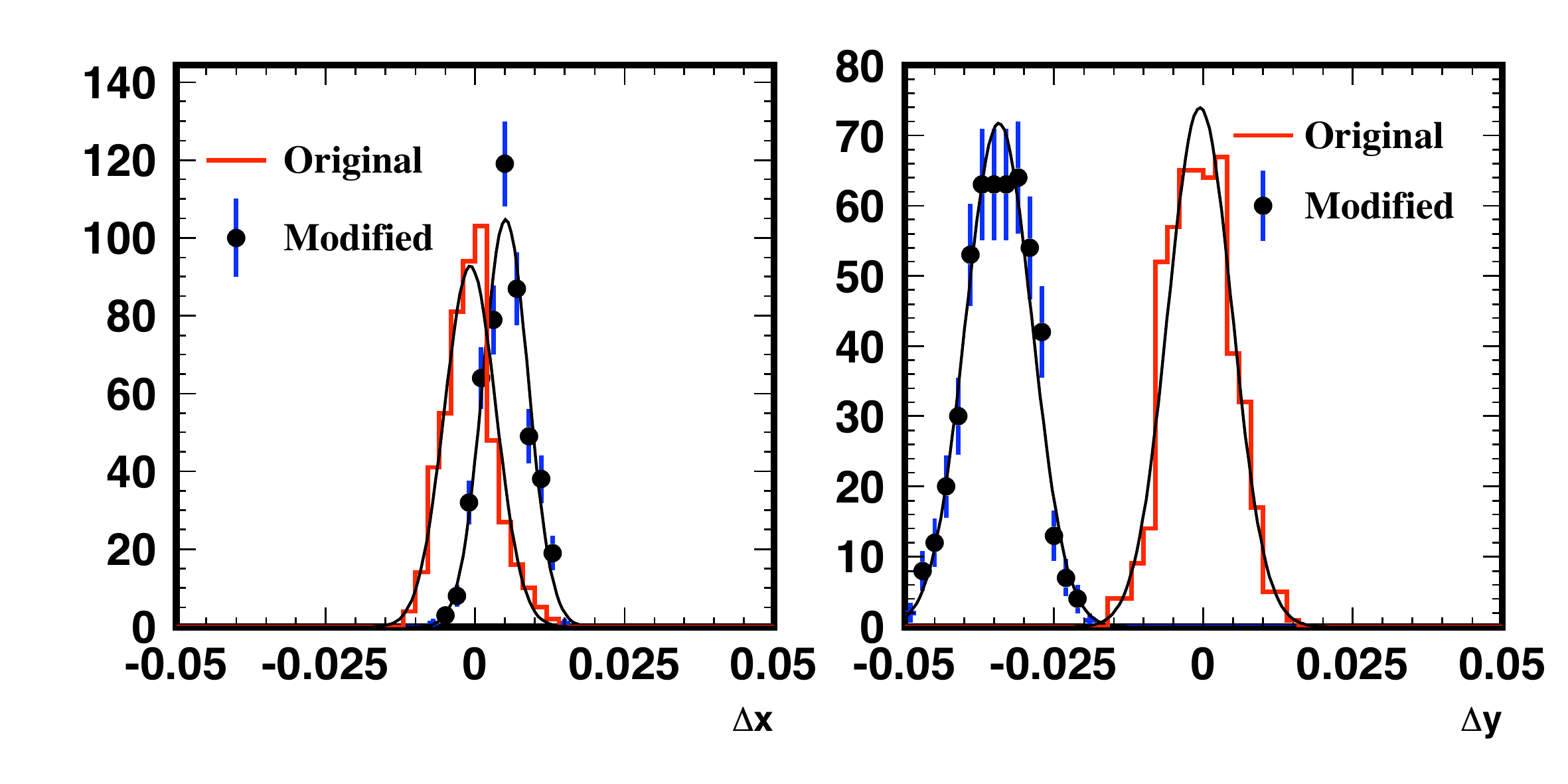}	\\[0.5ex]
  \includegraphics[width=7.5cm]{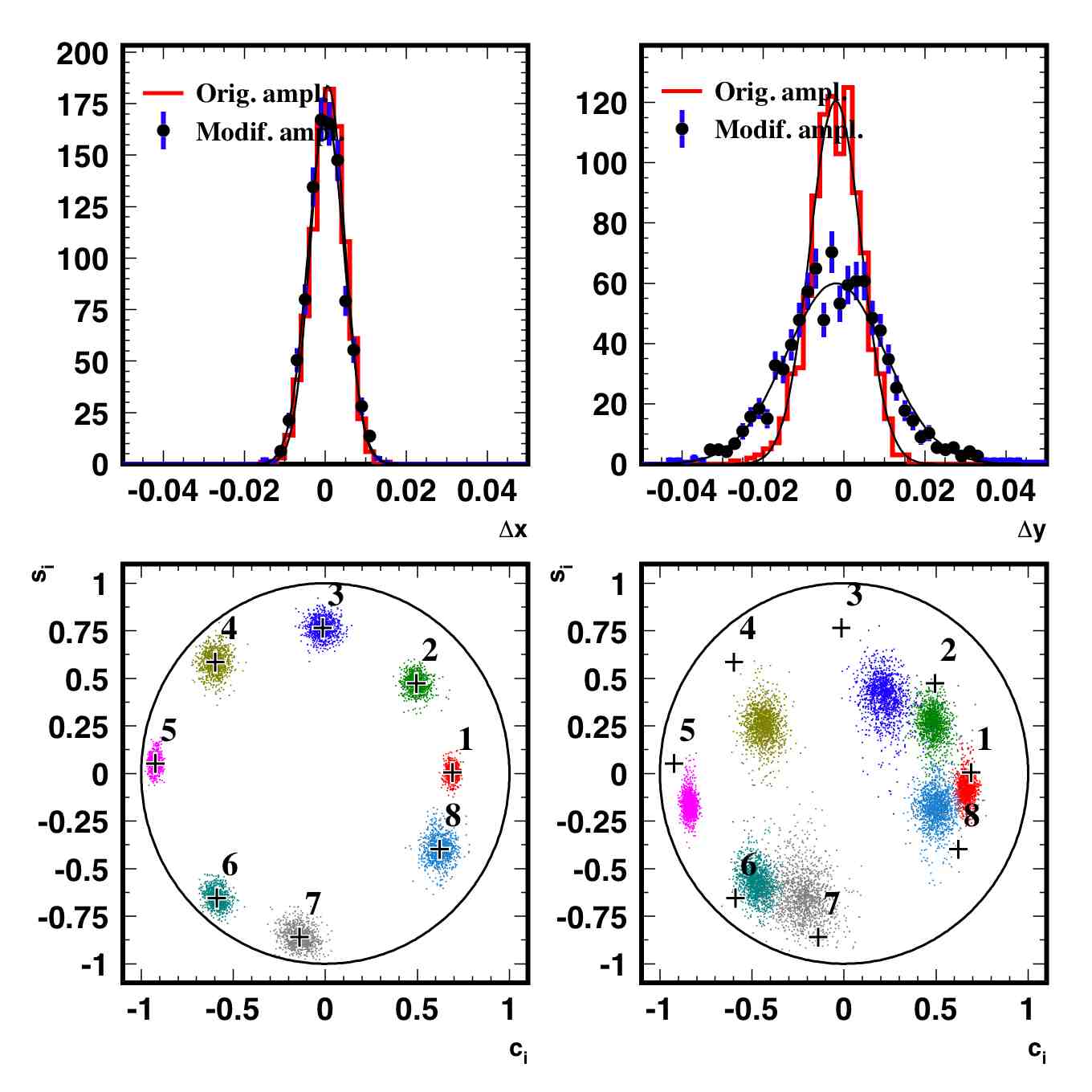}
  \caption{From~\protect\cite{bondar-poluektov-2008},
	showing results of toy Monte Carlo simulation of the $\phi_3/$Dalitz analysis.
	(Top:) using $\dmes_{CP}$ data;
	the histogram shows fit results where the same \dz\ decay amplitude
	is used for generation and binning, and
	the points with error bars show the case with different amplitudes.
	(Middle:) as above, but using
	$\psi(3770) \to (\ks\pi^+\pi^-)_{\dmes} (\ks\pi^+\pi^-)_{\dmes}$ data.
	(Bottom:) coefficients $(c_i,s_i)$ obtained from the eight
	$\left. \Delta \delta_D \right|_{model}$ bins 
	of Fig.~\protect\ref{fig:phi3-delta-binning},
	where (left) the same amplitude, and (right) different amplitudes
	are used for event generation and binning.}
  \label{fig:phi3-generation-and-binning}
\end{figure}

An advance has been the nontrivial binning shown in Fig.~\ref{fig:phi3-delta-binning},
which is unifotm in $\left. \Delta \delta_D \right|_{model}$.
However, results prove to be biased for finite $\dmes_{CP} \to \ks\pi^+\pi^-$
sample sizes: tests where events are generated in one model,
and reconstructed using another, find a return of model dependence,
as shown in Fig.~\ref{fig:phi3-generation-and-binning}.
It arises because for each bin, of the parameters in~(\ref{eq:c-s}),
only $c_i$ is reconstructed; $s_i$ is recovered by a $c_i^2 + s_i^2 = 1$ constraint.
A change of model results in a shift of the $\delta_D$ region sampled by a given bin,
introducing a bias in $s_i$ via the constraint.

The study~\cite{bondar-poluektov-2008} finds that if
$\{c_i,s_i\}$ are determined from
$\psi(3770) \to (\ks\pi\pi)_{\dmes} (\ks\pi\pi)_{\dmes}$ events,
the outcome is unbiassed for finite data samples:
a change of model can degrade the sensitivity
(Fig.~\ref{fig:phi3-generation-and-binning}), but not introduce a bias.
The importance of the entangled final state at the $\psi(3770)$ thus goes
beyond ``CP-tagging'':
the additional correlations in the final state
$(\ks\pi\pi)_{\dmes} (\ks\pi\pi)_{\dmes}$ are crucial.

\section{Summary}

Analyses which explicitly depend on the entanglement of neutral meson pairs
are becoming important in this field.
The effect allows the model-dependence in $\phi_3$/Dalitz analyses to be lifted:
CLEO-c data will already be necessary to enable full use to be made of
final \bmes-factory results;
BESIII results will be needed to exploit the data from a super-\bmes/flavor factory.
Entanglement in $\psi(3770) \to \dz\dzbar$ modulates tagged decay rates
in a way that must be taken into account for $\dz\to\kl\pi^0$ measurement,
and that has now enabled the first effective constraint,
$\delta_{\kmes\pi} = (22^{+11}_{-12} {}^{+9}_{-11})^\circ$,
on the strong phase difference in $\dz\to\kplus\pi^-$.
And at the \ufours, entanglement, used routinely in B-factory measurements,
has now been tested, even though a Bell inequality analysis cannot be performed.
A constraint on the decoherent fraction $\zeta_{\bz\bzbar} = 0.029 \pm 0.057$
is found; as for ``realistic'' local-realistic models,
the class obeying the Pompili-Selleri assumptions has been ruled out at $5.1\sigma$,
the first such constraint from data.


\section{Acknowledgements}
As well as expressing my appreciation to the FPCP 2008 conference organisers, 
I must thank my collaborators on Belle for discussions during the preparation
of the EPR correlations result: Apollo Go and Aurelio Bay (the principal authors),
Nick Hastings, Mike Peters, Samo Stani\v{c}, and others.
I would also like to acknowledge the contribution of colleagues at the
University of Queensland.
A preliminary version of the Belle results were presented at a colloquium there
in late 2006, and the optical analogy in Fig.~\ref{fig:optical-analogue} was
developed during stimulating discussions at that time.

\end{document}